\documentstyle[preprint,aps,epsf]{revtex}
\tightenlines
\begin{document}

\title{Probability of color singlet chain states in 
$e^+e^-$ annihilation}

\author{Qun Wang$^{1,2}$
\thanks{
qwang@public.sdu.edu.cn}, 
G\"osta Gustafson$^{1}$
\thanks{gosta@thep.lu.se}, 
Yi Jin$^{2}$
\thanks{tpc@sdu.edu.cn}
Qu-bing Xie$^{3}$ 
\thanks{xie@sdu.edu.cn}
}

\address{
$^1$ Department of Theoretical Physics, Lund University, 
S\"olvegatan 14A, S-22362 Lund, Sweden \\
$^2$ Physics Department, Shandong University, 
Jinan, Shandong 250100, P. R. China\\
$^3$ China Center of Advanced Science and Technology(World Lab)\\ 
P.O.Box 8730, Beijing 100080, P. R. China\\
}

\maketitle 

\begin{abstract}
We use the method of the color effective Hamiltonian to study 
the structure of color singlet chain states in $N_c=3$ 
and in the large $N_c$ limit. In order to obtain their total 
fraction when $N_c$ is finite, 
we illustrate how to orthogonalize these 
non-orthogonal states. We give numerical results 
for the fraction of orthogonalized states 
in $e^+e^-\rightarrow q\overline{q}gg$. 
With the help of a diagram technique, 
we derive their fraction up 
to $O(1/N_c^2)$ for the general multigluon process. 
For large $N_c$ the singlet chain states 
correspond to well-defined color
topologies. Therefore we may expect that the fraction of 
non-color-singlet-chain states is an estimate of 
the fraction of events where color 
reconnection is possible. 
In the case of soft gluon bremsstrahlung, 
we give an explicit form for 
the color effective Hamiltonian
which leads to the dipole 
cascade formulation for parton showering in leading order 
in $N_c$. The next-to-leading order 
corrections are also given  
for $e^+e^-\rightarrow q\overline{q}g_1g_2$ 
and $e^+e^-\rightarrow q\overline{q}g_1g_2g_3$.  

\end{abstract}

\begin{center}
(\today )
\end{center}

\section{Introduction}

Hadronic processes in various high energy collisions are generally described
in terms of two distinct phases: the perturbative phase and the
non-perturbative hadronization one. The perturbative phase is well described
by perturbative QCD (PQCD) while the hadronization phase cannot be described
from first principle and can only be described by phenomenological
models, e.g. the Lund string model\cite{and,sjo} or the cluster model\cite
{web}. These two phases are usually assumed to be well separated from
each other. It is believed that the cross section for the hadronic
process is fully determined by the perturbative phase, while in the
hadronization phase a definite hadron state is chosen with total probability
1. In both phases, however, the large $N_c$ (the number of color)
approximation is implied, which reduces the possible interference effects. A
color charge of one parton is specifically connected to its anti-color in
an accompanying parton, and with infinitely many colors the probability that
two (or more) partons have the same color is zero\cite{friberg}. So
here enters the phenomenological color flow method (CFM) commonly used in the
Lund model and the cluster model which implies 
assigning the color connection for
the final parton system\cite{sjo}. In these models, for every 
$e^{+}e^{-}\rightarrow q\overline{q}+ng$ event, a neutral color flow is
definitely determined and begins at the quark, connects each gluon one by
one in a certain order, and ends at the antiquark. Each flow piece spanned 
between two partons is color-neutral and its hadronization is treated in
a way similar to an independent $q\overline{q}$ 
singlet system. The present hadronization
models work well, which shows that the CFM or 
the large $N_c$ limit reflects some 
feature of the real world.

Recently we proposed a strict PQCD method to study the color structure of a
multiparton system. This method is called the method of color effective
Hamiltonian which is constructed from the PQCD invariant amplitude.
In this method a gluon is treated as an exact color octet, not a
bi-color or nonet. The goal of this paper is to
study the structure of the color singlet chain state 
(SCS or chain state) using this method for $N_c=3$ 
and in the large $N_c$ limit. 

The paper is organized as follows. In section \ref{h-c}, we 
outline the effective Hamiltonian method. 
In section \ref{sss-qqng}, we define SCS, and introduce a 
diagram method which is very efficient in 
calculating the color inner product of any two 
color states. In section \ref{ocesss}, we calculate 
the fraction of SCS for $N_c=3$. 
Since these states are not orthogonal to each other 
when $N_c$ is finite, we must find orthogonalized states 
to obtain a correct result. We introduce 
recipes for orthogonalization and use the 
orthogonalized states to calculate their fraction. 
As an example, we gives the numerical result 
for the fraction of orthogonalized SCS in the process 
of $e^+e^-$ annihilation into quark, antiquark and two gluons. 
Section \ref{cmesss-lnc} presents properties 
of SCS in the large $N_c$ limit. 
With the help of the diagram technique, 
we give their fraction up to $O(1/N_c^2)$ for general 
multigluon process. In section \ref{soft-gluon}, 
we give an explicit form for the momentum 
function $D$ in $H_c$ for the case of 
soft gluon bremsstrahlung. 
We show in the large $N_c$ limit 
the consistency of $H_c$ with the dipole cascade 
formulation for parton showering. 
We also discuss the next-to-leading order corrections 
for $e^+e^-\rightarrow q\overline{q}g_1g_2$ 
and $e^+e^-\rightarrow q\overline{q}g_1g_2g_3$.

\section{Color effective Hamiltonian}\label{h-c}

The QCD Lagrangian describes the $SU(3)_c$ gauge 
interaction of gluon fields 
$A_\mu ^\alpha $ $(\alpha =1,\cdots ,8)$ 
and quark fields $q$ with three colors $(R,Y,B)$. 
By redefining Gell-Mann Matrices and gluon fields, 
we obtain eight new fields ($X_\mu ^{12}$, 
$X_\mu ^{21}$, $X_\mu ^{23}$, $X_\mu ^{32}$, 
$X_\mu ^{31}$, $X_\mu ^{13}$, $X_\mu ^{3}$, 
$X_\mu ^{8}$) which couple to color combinations 
($Y\overline{R}$, $R\overline{Y}$, $B\overline{Y}$, 
$Y\overline{B}$, $R\overline{B}$, $B\overline{R}$, 
$R\overline{R}/Y\overline{Y}$, 
$R\overline{R}/Y\overline{Y}/B\overline{B}$) 
respectively. Hence we write the QCD Lagrangian 
in a form with the quark-gluon interaction term 
showing clear color significance. 
This provides a 
strict formulation for calculating the fraction 
of color singlets for a multiparton system 
at the tree level from PQCD\cite{wang,wang99}. For the process 
$e^{+}e^{-}\rightarrow q\overline{q}+ng$, the
essential part of the formulation is to 
exploit the color effective 
Hamiltonian $H_c$ to compute the amplitude 
$\left\langle f\right| H_c\left|0\right\rangle $ 
for a certain color state $\left| f\right\rangle $. 
The color effective Hamiltonian $H_c$ is found 
from the invariant amplitude 
$M_{ab}^{\alpha _1\alpha _2\cdots \alpha _n}$ to be: 
\begin{equation} 
\label{hc0} 
\begin{array}{c} 
H_c=\sum_P(T^{\alpha _{P(1)}}T^{\alpha _{P(2)}}\cdots T^{\alpha 
_{P(n)}})_{ab}D^P\Psi _a^{\dagger }\Psi ^{b\dagger }A_1^{\alpha _1\dagger
}A_2^{\alpha _2\dagger }\cdots A_n^{\alpha _n\dagger } \\ 
=\sum_P(1/\sqrt{2})^nTr(Q^{\dagger }G_{P(1)}^{\dagger }G_{P(2)}^{\dagger
}\cdots G_{P(n)}^{\dagger })D^P\qquad \qquad \quad 
\end{array}
\end{equation}
where $\Psi _a^{\dagger }$ and 
$\Psi ^{b\dagger }$ 
are color operators for the quark and antiquark; 
$(Q^{\dagger })_a^b=\Psi ^{b\dagger }\Psi _a^{\dagger }$ 
is the nonet tensor operator; $G_u^{\dagger }$ is 
the gluon's color octet operator; 
$D^P$ is a function of parton momenta and 
its dependence on the order of partons 
is marked by $P$ which denotes a permutation of 
parton labels $(1,2,\ldots,n)$. 
The color effective Hamiltonian is another expression for the $S$ matrix, 
and is therefore not Hermitian. One can verify its validity by 
using $H_c$ to calculate the matrix element for 
the process $e^{+}e^{-}\rightarrow q \overline{q}+ng$, which 
returns the original 
$|M_{ab}^{\alpha _1\alpha _2\cdots \alpha _n}|^2$. 
In order to make $H_c$ useful, we can define color 
states for the parton system $q \overline{q}+ng$ independently. 
A color state can be defined in the color 
space where each gluon subspace is either 
$\underline{8}$ or a little larger 
$\underline{3}\otimes \underline{3}^{*}$. 
The effects of the unphysical ''singlet-gluon'' state, 
brought by enlarging the 
color space $\underline{8}$ to 
$\underline{3}\otimes \underline{3}^{*}
(=\underline{1}\oplus \underline{8})$, can be 
eliminated by calculating the projection of the 
state on $H_c$.  
There are many ways of reducing the 
color space. Corresponding to each reduction recepe there 
is one set of orthogonal singlets spaces 
whose bases make up a complete and orthogonal 
set of color singlets. $H_c$ has the property 
from unitarity, that the sum of the cross sections over 
all color singlets in a complete and orthogonal 
singlet set for the system $q \overline{q}+ng$, is 
equal to the total cross section of 
$e^+e^-\rightarrow q\overline{q}+ng$ at 
the tree level.

\section{Singlet chain states 
in $q\overline{q}+ng$ system}\label{sss-qqng}

A chain state for a $q\overline{q}+ng$ system 
is made up of a chain of 
$(n+1)$ pieces. Each piece is a singlet 
formed by the color charge of one parton and its 
anti-charge of the other one, and its color 
structure is the same as a $q\overline{q}$ system. 
A neutral color flow is also composed of pieces, 
each of which is formed by the color charge of one parton 
and its anti-charge of the other one, but it is 
a neutral color state not necessarily a color singlet. 
A singlet chain state can be decomposed 
into neutral color flow states. 
There are $n!$ chain states which connect $q$ 
via $n$ gluons in a specified order to $\overline{q}$, where 
the order of gluons is denoted by a permutation 
of $(1,2,\cdots ,n)$: $P(1),P(2),\cdots ,P(n)$. 
These states can be written as

\begin{equation}
\label{singlet} 
(\left| f_i\right\rangle ,i=1,2,\cdots ,n!)
\equiv \{ N_c^{-(n+1)/2}\left| 
1_{qP(1)}1_{P(1)P(2)}1_{P(2)P(3)}
\cdots 1_{P(n)\overline{q}}\right\rangle   \}  
\end{equation}
where $N_c^{-(n+1)/2}$ is a normalization factor. 
Any two different chain states,
$\left| f_i\right\rangle $ and $\left|
f_j\right\rangle $, are not orthogonal to each other, i.e. 
$\langle f_i\mid f_j\rangle \neq 0$.
However, we can prove that any two
different chain states are 
approximately orthogonal to each other
to order $1/N_c^2$. As an example, 
we look at the inner product of
two states as follows:
$$
\begin{array}{l}
\left| f_1\right\rangle = N_c^{-(n+1)/2}\left|
1_{q1}1_{12}1_{23}1_{34}1_{45}\cdots 1_{i,i+1}\cdots 1_{n 
\overline{q}}\right\rangle \\ 
\left| f_2\right\rangle =
N_c^{-(n+1)/2}\left| 1_{q1}1_{13}1_{32}1_{24}1_{45}\cdots 1_{i,i+1}\cdots
1_{n\overline{q}}\right\rangle \\
C_{12}\equiv \langle f_1\mid f_2\rangle \\
=N_c^{-(n+1)}\langle
1_{q1}1_{12}1_{23}1_{34}1_{45}\cdots 1_{i,i+1}\cdots 1_{n\overline{q}}\mid 
1_{q1}1_{13}1_{32}1_{24}1_{45}\cdots 1_{i,i+1}\cdots 1_{n\overline{q}
}\rangle
\end{array}
$$
The gluon order in $\left| f_1\right\rangle $ and 
$\left| f_2\right\rangle $ is $(12345\cdots ii+1\cdots n)$ and $(13245\cdots 
ii+1\cdots n)$ respectively. The only difference is that the position of 
$g_2 $ and $g_3$ is interchanged in two states. We write the explicit form 
of $\left| f_1\right\rangle $ and $\left| f_2\right\rangle $ as: 
$$ 
\begin{array}{l} 
\left| f_1\right\rangle =N_c^{-(n+1)/2}\delta (a_0,b_1)\delta
(a_1,b_2)\delta (a_2,b_3)\delta (a_3,b_4)\delta (a_4,b_5)\cdots \delta
(a_i,b_{i+1})\cdots \delta (a_n,b_0) \\ 
\left| \Psi _{a_0}\Psi ^{b_1}\Psi _{a_1}\Psi ^{b_2}\Psi _{a_2}\Psi
^{b_3}\Psi _{a_3}\Psi ^{b_4}\Psi _{a_4}\Psi ^{b_5}\cdots \Psi _{a_i}\Psi
^{b_{i+1}}\cdots \Psi _{a_n}\Psi ^{b_0}\right\rangle 
\end{array}
$$
$$
\begin{array}{l}
\left| f_2\right\rangle =N_c^{-(n+1)/2}\delta (a_0^{\prime },b_1^{\prime
})\delta (a_1^{\prime },b_3^{\prime })\delta (a_3^{\prime },b_2^{\prime
})\delta (a_2^{\prime },b_4^{\prime })\delta (a_4^{\prime },b_5^{\prime
})\cdots \delta (a_i^{\prime },b_{i+1}^{\prime })\cdots \delta (a_n^{\prime
},b_0^{\prime }) \\ 
\left| \Psi _{a_0^{\prime }}\Psi ^{b_1^{\prime }}\Psi _{a_1^{\prime }}\Psi
^{b_3^{\prime }}\Psi _{a_3^{\prime }}\Psi ^{b_2^{\prime }}\Psi _{a_2^{\prime
}}\Psi ^{b_4^{\prime }}\Psi _{a_4^{\prime }}\Psi ^{b_5^{\prime }}\cdots \Psi
_{a_i^{\prime}}\Psi ^{b_{i+1}^{\prime}}
\cdots \Psi _{a_n}\Psi ^{b_0}\right\rangle 
\end{array}
$$
where $a$,$b$,$a^{\prime }$,$b^{\prime }$ are the 
color and anticolor indices, and 
$0,1,2,3,\cdots ,n$ denote $q(\overline{q})$, 
$g_1$, $g_2$, $g_3$,$\cdots 
$,$g_n$. Thus $\langle f_1\mid f_2\rangle $ is: 
\begin{equation}
\label{inner1} 
\begin{array}{l}
\langle f_1\mid f_2\rangle =N_c^{-(n+1)}[\delta (a_0,b_1)\delta
(a_1,b_2)\delta (a_2,b_3)\delta (a_3,b_4)\delta (a_4,b_5)\cdots \delta
(a_i,b_{i+1})\cdots \delta (a_n,b_0)] \\ 
\cdot [\delta (a_0^{\prime },b_1^{\prime })\delta (a_1^{\prime },b_3^{\prime
})\delta (a_3^{\prime },b_2^{\prime })\delta (a_2^{\prime },b_4^{\prime
})\delta (a_4^{\prime },b_5^{\prime })\cdots \delta (a_i^{\prime
},b_{i+1}^{\prime })\cdots \delta (a_n^{\prime },b_0^{\prime })]\\
\cdot \prod_{u=0,1,\cdots ,n}\delta (a_u,a_u^{\prime })
\delta (b_u,b_u^{\prime }) 
\end{array}
\end{equation}
When $\prod_{u=0,1,\cdots ,n}\delta (a_u,a_u^{\prime })\delta
(b_u,b_u^{\prime })$ is contracted with the 
content in the second square 
bracket, the above equation becomes 
\begin{equation}
\label{inner2} 
\begin{array}{l}
\langle f_1\mid f_2\rangle =N_c^{-(n+1)}[\delta (a_0,b_1)\delta
(a_1,b_2)\delta (a_2,b_3)\delta (a_3,b_4)\delta (a_4,b_5)\cdots \delta
(a_i,b_{i+1})\cdots \delta (a_n,b_0)] \\ 
\cdot [\delta (a_0,b_1)\delta (a_1,b_3)\delta (a_3,b_2)\delta
(a_2,b_4)\delta (a_4,b_5)\cdots \delta (a_i,b_{i+1})\cdots \delta (a_n,b_0)]
\\ 
=N_c^{-(n+1)}\cdot N_c\cdot N_c^{n-2}=1/N_c^2 
\end{array}
\end{equation}

We can use diagrams to visualize and 
simplify our calculation of inner
products. Let us write the color and 
anticolor indices into two rows where
the numbers in the first one are anticolor 
indices and those in the
second are color ones. We draw a line between 
the number $v$ in the first
row and the number $u$ in the second row if 
there is a $\delta ($$\,a_u,b_v)$
in $\langle f_i\mid f_j\rangle $. As a rule, 
we have $\langle f_i\mid
f_j\rangle =N_c^{l-n-1}$ where $l$ is the 
number of closed paths. We can
verify eq.(\ref{inner2}) by drawing the 
corresponding diagram. As an example
of (\ref{inner2}), we show the $l=2$ case in Fig.1.

Generally, we can carry out the inner 
product of any two states by a diagram:
\begin{equation}
\label{inner3} 
\begin{array}{l}
\langle f_i\mid f_j\rangle \\
=N_c^{-(n+1)}[\delta (a_0,b_{P\{1)})\delta
(a_{P(1)},b_{P(2)})\delta (a_{P(2)},b_{P(3)})\cdots \delta
(a_{P(i)},b_{P(i+1)})\cdots \delta (a_{P(n)},b_0)] \\ 
\cdot [\delta (a_0,b_{P^{\prime }\{1)})\delta (a_{P^{\prime
}(1)},b_{P^{\prime }(2)})\delta (a_{P^{\prime }(2)},b_{P^{\prime
}(3)})\cdots \delta (a_{P^{\prime }(i)},b_{P^{\prime }(i+1)})\cdots \delta
(a_{P^{\prime }(n)},b_0)] \\ 
=N_c^{-(n+1)}\cdot N_c^{n_1}\cdot N_c^{n_2}=N_c^{n_1+n_2-(n+1)} 
\end{array}
\end{equation}
If there is a factor $\delta (a_{P(i)},b_{P(i+1)})\delta (a_{P^{\prime 
}(j)},b_{P^{\prime }(j+1)})$ where $P(i)=P^{\prime }(j)$ and 
$P(i+1)=P^{\prime }(j+1)$, we get one factor $N_c$. 
$n_2$ denotes the number of 
such identical $\delta$-symbols in two square brackets of 
(\ref{inner3}). If we exclude the 
identical $\delta$-symbols in the two brackets, the rest contains 
only different $\delta $s. 
We denote the number of such $\delta $s in one 
bracket as $n_d$. It can be demonstrated from the 
diagram that when $n_d$ is odd, these different $\delta $s 
form one loop, and when $n_d$ is even they 
form two loops. $n_1$ denotes the number of loops 
formed by those different $\delta $s and is 2/1 for 
even/odd $n_d$. The maximum of $n_1+n_2-(n+1)$ is $-2$, which 
occurs when $n_2$ reaches its 
maximum value $n-2$. In this case the only 
difference between the two states 
is that only two gluons have their positions 
interchanged, just what we see in (\ref{inner1}) 
and (\ref{inner2}). Hence we see that 
any two different chain states are 
approximately orthogonal to each other 
to the order of $1/N_c^2$.

\section{Orthogonalization and fraction of 
singlet chain states at finite $N_c$ }
\label{ocesss}

For finite $N_c$, as in Nature, the chain states 
in (\ref{singlet}) are not 
orthogonal to each other. We cannot 
directly take the sum of each 
$\left| \left\langle f_i\right|H_c\left| 0\right\rangle \right| ^2$ 
to give the total fraction for chain 
states, because it would make the contribution from the 
overlapped part of any two different states counted multiply. 
In this section we find a set of orthogonal 
states based on $\{|f_i\rangle \}$. 

For the process $e^+e^-\rightarrow q\overline{q}g_1\cdots g_n$, 
there are $n!$ chain states $\{|f_i\rangle \}$, each 
connecting $q$ with $\overline{q}$  
through $n$ gluons in a specific order. 
A set of orthogonal states $\{|f^{1}_i\rangle \}$ is 
related to $\{|f_j\rangle \}$ 
by a linear transformation $U^{1}_{ij}$, i.e. 
$|f^{1}_i\rangle =U^{1}_{ij}|f_j\rangle $, where 
$U^{1}_{ij}$ is not a unitary matrix because of 
the non-orthogonality of the original states $\{|f_j\rangle \}$. 
$U^{1}_{ij}$ is not unique, so one can find 
many different ways of orthogonalization. 
Two different matrices $U^{1}$ and $U^{2}$ are associated 
with each other via a unitary matrix $U^{12}$, which guarantees 
the conservation of probability. This means that for 
two sets of orthogonal states, $|f^{1}_i\rangle $ and 
$|f^{2}_i\rangle $, which are both associated with  
the same set of non-orthogonal states $\{|f_j\rangle \}$, the 
following identity holds: 
$\sum _i \left| \left\langle f^{1}_i| H_c \right\rangle
\right| ^2=\sum _i\left| \left\langle f^{2}_i| H_c \right\rangle \right|^2$. 

Due the rapidly growing complexity of orthogonalization 
for large numbers of emitted gluons, let us only discuss 
the two simplest cases $e^+e^-\rightarrow q\overline{q}g_1g_2$ 
and $e^+e^-\rightarrow q\overline{q}g_1g_2g_3$. 

For $e^+e^-\rightarrow q\overline{q}g_1g_2$, 
there are two chain states: 
$$(\left| f_1\right\rangle ,\left| f_2\right\rangle )\equiv 
(\frac 1{3\sqrt{3}}\left| 1_{q1}1_{12}1_{2\overline{q}}
\right\rangle ,\frac 1{3\sqrt{3}}\left| 1_{q2}1_{21}
1_{1\overline{q}}\right\rangle )$$
where $\left|f_1\right\rangle $ and $\left| f_2\right\rangle $ 
are not orthogonal to each other: 
$\langle f_1\mid f_2\rangle =1/N_c^2=1/9$. 
A straightforward way to construct orthogonal states is to 
linearly transform $\left| f_1\right\rangle $ and 
$\left|f_2\right\rangle $ into symmetric and anti-symmetric 
states respectively, i.e.
\begin{equation} 
\label{sys-st}
\left( 
\begin{array}{c}
\left| f_1'\right\rangle \\ 
\left| f_2'\right\rangle 
\end{array} 
\right) 
=\left( 
\begin{array}{cc}
\frac 3{\sqrt{20}} & \frac 3{
\sqrt{20}} \\ \frac 34 & -\frac 34 
\end{array}
\right) 
\left( 
\begin{array}{c}
\left| f_1\right\rangle \\ 
\left| f_2\right\rangle 
\end{array}
\right), 
\end{equation}
where $\left| f_1' \right\rangle $ is the symmetric and 
$\left| f_2' \right\rangle $ the anti-symmetric state. 
Using $\left\langle f_1 | H_c \right\rangle =  
\frac{32}{9\sqrt{3}}D^{12}-\frac 4{9\sqrt{3}}D^{21}$ and 
$\left\langle f_2 | H_c \right\rangle =
-\frac 4{9\sqrt{3}}D^{12}+\frac{32}{9\sqrt{3}}D^{21}$,  
we obtain the sum of squared projections as follows: 
\begin{equation}
\label{prob1} 
\begin{array}{l}
\left| \left\langle f_1'| H_c \right\rangle \right|^2+
\left| \left\langle f_2'| H_c \right\rangle \right|^2 \\ 
=(\frac{14^2}{27\cdot 5}+3)(\left| D^{12}\right| ^2+\left| D^{21}\right| 
^2)+2( \frac{14^2}{27\cdot 5}-3)Re(D^{12}\cdot D^{21*}) \\ 
\approx 4.45(\left| D^{12}\right| ^2+\left| D^{21}\right| ^2)
-3.1Re(D^{12}\cdot D^{21*})
\end{array}
\end{equation}
The invariant amplitude for 
$e^{+}e^{-}\rightarrow q\overline{q}g_1g_2$ is:
$M_{ab}^{\alpha _1\alpha _2}=(T^{\alpha _1}T^{\alpha
_2})_{ab}D^{12}+(T^{\alpha _2}T^{\alpha _1})_{ab}D^{21}$.
The total cross section of 
$e^{+}e^{-}\rightarrow q\overline{q}g_1g_2$ at 
the tree level is then: 
\begin{equation}
\label{cs-tr-2g} 
\begin{array}{c}
\sigma _{tree}(e^{+}e^{-}\rightarrow q 
\overline{q}g_1g_2)=\int d\Omega \; M_{ab}^{\alpha _1\alpha _2}
(M_{ab}^{\alpha _1\alpha _2})^*\\
=\int d\Omega \;[\frac{16}3(\left| D^{12}\right| ^2+\left| D^{21}\right| ^2)-
\frac43Re(D^{12}\cdot D^{21*})] 
\end{array}
\end{equation}
The functions $D^{12}$ and $D^{21}$ correspond to different kinematical 
distributions, which in general have a rather limited overlap. Therefore
the kinematic interference term proportional 
to $Re(D^{12} \cdot D^{21*})$ is 
suppressed. This kinematic interference 
term can be calculated in 2nd order
perturbation theory (we will come to this later). 
The result depends on the kinematical configuration,
e. g. expressed by a $y$-cut for the definition of the 4-jet events. 
If we neglect kinematic interference terms in (\ref{prob1}) 
and (\ref{cs-tr-2g}), we obtain: 
\begin{equation}
\label{prob2}
P(e^{+}e^{-}\rightarrow q\overline{q}g_1g_2\rightarrow SCS\,)=\frac{\sigma
(e^{+}e^{-}\rightarrow q\overline{q}g_1g_2\rightarrow SCS\,)}
{\sigma _{tree}(e^{+}e^{-}
\rightarrow q\overline{q}g_1g_2)}\simeq 83\%
\end{equation}
where we mention again that SCS is the abbreviation 
for singlet chain states. 

In order to evaluate the approximation of dropping 
kinematic interference terms, let us calculate the 
fraction $P(e^{+}e^{-}\rightarrow
q\overline{q}g_1g_2\rightarrow SCS\,)$ exactly. 
For a higher order process 
with more gluons produced, the calculation is too complicated. 
We know there are 8 lowest order Feynman diagrams in 
$e^{+}e^{-}\rightarrow q\overline{q}g_1g_2$, containing 2 diagrams with 
a tri-gluon vertex. In the calculation 
we choose the Feynman gauge, and we replace the polarization sum 
$\sum _{\lambda =1,2}\epsilon ^{\mu}_{\lambda}(p)
\epsilon ^{\nu}_{\lambda}(p)$ with $-g^{\mu\nu}$, where 
the sum is taken over two transverse polarizations 
$\lambda =1,2$. However, $-g^{\mu\nu}$ equals to the sum over 
all four polarizations, including two unphysical ones. 
To cancel unphysical polarization states 
and guarantee unitarity, 
we should introduce two ghost diagrams. 
Of course, we may work in the physical 
gauge, where there is no ghost, and directly use physical 
polarizations for the gluons, but the calculation 
in the Feynman gauge is much simpler; 
see Ref.\cite{peskin} for details. 
Thus, including two ghost diagrams, 
we have 10 diagrams altogether. 
The fact that ghost diagrams don't interfere with the 8 gluon 
diagrams make them easier to deal with. 
When we calculate the square of the amplitude 
$M=M_1+M_2+...+M_{10}$, we know that 
a non-interference term $|M_i|^2$ has a color factor $16/3$, 
while an interference term $M_iM_j^*$ has a color 
factor $-2/3$. For a tri-gluon and ghost-ghost-gluon vertex, 
we make substitution $T^aT^b-T^bT^a=if^{abc}T^c$ to get the functions 
$D^{12}$ and $D^{21}$. According to 
Eq.(\ref{prob1}) and (\ref{cs-tr-2g}), we have: 
\begin{equation}
\label{prob3}
P(e^{+}e^{-}\rightarrow q\overline{q}g_1g_2\rightarrow SCS\,)
=\frac{\int d\Omega \;[4.45(\left| D^{12}\right| ^2+\left| D^{21}\right| ^2)
-3.1Re(D^{12}\cdot D^{21*})]}
{\int d\Omega \;[\frac{16}3(\left| D^{12}\right| ^2+\left| D^{21}\right| ^2)
-\frac43Re(D^{12}\cdot D^{21*})] }
\end{equation}
We use a Monte Carlo method to do the 
phase space integration and evaluate this ratio. 
The result is shown in Fig.4 as a function of a cutoff 
$y_{cut}=(p_i+p_j)\cdot (p_i+p_j)/s$ where $p_i$ is 
the 4-momentum of parton $i$ and $\sqrt{s}$ is the center-of-mass energy of 
the $e^{+}e^{-}$ collision, which we set to 91GeV. In the figure, we see 
that the rate decreases slowly, from 0.72 to 0.67, as $y_{cut}$ varies 
from $10^{-4}$ to $10^{-2}$. These values are smaller than the 
rate 0.83 obtained by neglecting the kinematic 
interference contribution, but the difference is not large. 
This implies that kinematic interference terms are less 
important than the non-interference terms.

For the $q\overline{q}g_1g_2g_3$ system, there are 6 singlet chain 
states $( \left| f_i\right\rangle ,i=1,2\cdots 6 )$ which connect 
$q$ to $\overline{q}$ via three ordered gluons: 
123, 231, 312, 213, 132, 321, respectively. 
These states are not orthogonal to each other. 
The inner product $\langle f_i\mid f_j\rangle $ is 
1 for $i=j$ and 1/9 for $i\neq j$.  
Our goal is to find 6 orthogonal 
states from them. The new states are denoted 
$(\left| f_i'\right\rangle ,\,\,i=1,2\cdots 6)$. 
Making use of the symmetric and approximately orthogonal 
properties of $\langle f_i\mid f_j\rangle $, we find one 
orthogonal set as follows:
\begin{equation}
\left| f_i'\right\rangle =(1+\sigma )\left|
f_i\right\rangle +\epsilon \sum_{j\neq i}\left| f_j\right\rangle
\,,\,\,\,\,\,\,\,for\,\,i=1,\cdots ,6 
\end{equation}
where $\sigma \approx 0.021$, $\epsilon \approx -0.037$. 
According to arguments given for 
$e^+e^-\rightarrow q\overline{q}g_1g_2$, we may find a 
different set of orthogonal states which is related to 
$\left| f_i'\right\rangle $ via a unitary transformation, 
and we know that either set is equivalent.

Now we try to calculate the total fraction for chain states 
$\left| f_i^{\prime}\right\rangle $ by projecting them 
on $|H_c\rangle $. Since each state is orthogonal 
to any other, we can sum up all squared projections: 
\begin{equation}
\label{sum-fhc} 
\begin{array}{l}
\sum_{i=1}^6\left| \langle f_i^{\prime }\mid H_c\rangle \right|
^2=5.47\{\left| D^{123}\right| ^2+\left| D^{231}\right| ^2+\left|
D^{312}\right| ^2 \\ 
+\left| D^{213}\right| ^2+\left| D^{132}\right| ^2
+\left| D^{321}\right|^2\}\\
+kinematic\,\,interference\,\,terms 
\end{array}
\end{equation}
where kinematic interference terms 
refer to interference terms between 
two different $D$. 
Thus we obtain for chain states in 
$e^{+}e^{-}\rightarrow q\overline{q}g_1g_2g_3$: 
\begin{equation}
\label{c-sec1}
\sigma (e^{+}e^{-}\rightarrow 
SCS)=\int d\Omega \sum_{i=1}^6
\left| \langle f_i'\mid H_c\rangle \right| ^2 
\end{equation}
In order to estimate the fraction of chain states, we need to 
know the total cross section for $e^{+}e^{-}\rightarrow q\overline{q}
g_1g_2g_3 $ at the tree level: 
\begin{equation}
\label{c-sec2} 
\begin{array}{l}
\sigma _{tree}(e^{+}e^{-}\rightarrow q 
\overline{q}g_1g_2g_3)=\int d\Omega \cdot M_{ab}^{\alpha
_1\alpha _2\alpha _3}\cdot (M_{ab}^{\alpha _1\alpha _2\alpha _3})^{*} \\ 
=\int d\Omega 
(\sum_{P}(T^{\alpha _{P(1)}}T^{\alpha _{P(2)}}T^{\alpha _{P(3)}})_{ab}D^P)
(\sum_{P'}(T^{\alpha _{P'(1)}}T^{\alpha _{P'(2)}}T^{\alpha _{P'(3)}})_{ab} 
D^{P'})^{*} 
\end{array}
\end{equation}
Here we resume the convention that a repetition 
of indices stands for summation. Expanding (\ref{c-sec2}) gives:
\begin{equation}
\label{c-sec3}
\begin{array}{l}
\sigma _{tree}(e^{+}e^{-}\rightarrow q 
\overline{q}g_1g_2g_3) \\ 
=\int d\Omega \,\cdot \{\frac 12 
\frac{(N_c^2-1)^3}{(2N_c)^2}[\left| D^{123}\right| ^2+\left| D^{231}\right|
^2+\left| D^{312}\right| ^2 \\ 
+\left| D^{213}\right| ^2+\left|
D^{132}\right| ^2+\left| D^{321}\right| ^2]\\
+kinematic\,\,interference\;\;terms \} 
\end{array}
\end{equation}
where kinematic interference terms 
are suppressed by powers of $1/N_c$. Thus, to 
leading order in $N_c$, we can give an instant estimate for 
the fraction from (\ref{c-sec1}) and (\ref{c-sec3}) without 
carrying out phase space integrals: 
$$
P(e^{+}e^{-}\rightarrow SCS)=\frac{\sigma (e^{+}e^{-}\rightarrow SCS)}
{\sigma _{tree}(e^{+}e^{-}\rightarrow q\overline{q}g_1g_2g_3)}
=\frac{5.47}{\frac 12\frac{(N_c^2-1)^3}{(2N_c)^2}}\sim 77\% 
$$

In this section, we have discussed the orthogonalization 
for chain states and estimated their fraction 
for $e^+e^-\rightarrow q\overline{q}g_1g_2$ and 
$e^+e^-\rightarrow q\overline{q}g_1g_2g_3$ by neglecting 
kinematic interference and then keeping only the 
interference due to the finite number of colors, $N_c$. 
There are many ways of constructing orthogonal 
states from the original non-orthogonal ones. 
Different ways lead to different orthogonal states. 
They are, however, equivalent for calculations of probabilities. 
Normally we can make use of the fact that the original states 
are approximately orthogonal up to $O(1/N_c^2)$. 
Thus one can find a set of orthogonal 
states, which are slightly different 
from the original ones, i.e. the transformation 
matrix is close to the unit matrix. 
The other straightforward orthogonalization 
recipe we give in this section, is to symmetrize 
and anti-symmetrize non-orthogonal states. 
We know that kinematic interference terms are 
all suppressed by $O(1/N_c)$ with respect to 
the non-interference terms. 
If we neglect all kinematic interference 
terms and then keep only the color 
interference brought by finite $N_c$, 
the total fraction is 83\% 
for $e^+e^-\rightarrow q\overline{q}g_1g_2$ and 
$77\%$ for $e^+e^-\rightarrow q\overline{q}g_1g_2g_3$. 
For the sake of estimating the magnitude of 
the kinematic interference, we give the numerical result 
for the fraction $P(e^{+}e^{-}\rightarrow
q\overline{q}g_1g_2\rightarrow SCS\,)$ 
with the kinematic interference taken into account. 
The result is shown in Fig.4 as a function of 
$y_{cut}$. We see that the rate decreases slowly, 
from 0.72 to 0.67, as $y_{cut}$ varies 
from $10^{-4}$ to $10^{-2}$. These values are smaller than the 
rate 0.83 obtained by neglecting kinematic interference terms, 
but the difference is not large. This implies that 
kinematic interference terms are less important than 
non-interference ones, though not negligibly small.

\section{Properties of singlet chain states for 
$q\overline{q}+ng$ in the 
large $N_c$ limit}
\label{cmesss-lnc}

In this section, we will study the properties 
of the chain states
for $q\overline{q}+ng$ 
in the large $N_c$ limit and obtain their 
fraction to $O(1/N_c^2)$. 

The projection of a chain state 
$\left| f\right\rangle $ on $|H_c\rangle $ is: 
$$
\langle f\mid H_c\rangle 
=\sum_P(1/\sqrt{2})^nD^P\langle 
f\mid Tr(QG_{P(1)}G_{P(2)}\cdots G_{P(n)})\rangle 
$$
where $H_c$ is given by (\ref{hc0}). Without loss of 
generality, we choose 
$$\left| f\right\rangle  
=N_c^{-(n+1)/2}\left|
1_{q1}1_{12}1_{23}\cdots 1_{i,i+1}
\cdots 1_{n\overline{q}}\right\rangle $$
The projection of any other chain state can be obtained 
by permuting gluon labels.
For convenience, we will ignore the 
normalization factor $N_c^{-(n+1)/2}$ and 
denote $|f\rangle $ as equivalent to 
$\left| 1_{q1}1_{12}1_{23}\cdots 1_{i,i+1}\cdots 1_{n\overline{q}}
\right\rangle $ in intermediate steps of the calculation. We will 
put the normalization factor back in
the final results. 

The order of the
$n$ gluons in $\left| f\right\rangle $ is 
$(1,2,3,\cdots ,n)$. In $|H_c\rangle $, 
there is also a term $|Tr(QG_1G_2\cdots
G_n)\rangle $ with gluon labels in the same order. 
Let us first calculate 
\begin{equation}
\label{ftrq}
\langle f\mid Tr(QG_1G_2\cdots G_n)\rangle =N_c^{-(n+1)/2}\langle
1_{q1}1_{12}1_{23}\cdots 1_{i,i+1}\cdots 1_{n\overline{q}}\mid
Tr(QG_1G_2\cdots G_n)\rangle  
\end{equation}
We can expand $\left| Tr(QG_1G_2\cdots G_n)\right\rangle $ as follows: 
\begin{equation}
\label{trq}
\begin{array}{l}
\left| Tr(QG_1G_2\cdots G_n)\right\rangle =\left| Tr(QG_1^{\prime
}G_2^{\prime }\cdots G_n^{\prime })\right\rangle  \\ 
+\sum_{k=1}^n\sum_{\{u_1,\cdots,u_k\}}
(-\frac 1{N_c})^k\left| 1_{u_1}1_{u_2}\cdots
1_{u_k}\right\rangle \left| Tr(QG_{v_1}^{\prime }G_{v_2}^{\prime }\cdots
G_{v_{n-k}}^{\prime })\right\rangle 
\end{array}
\end{equation}
where $(v_1,v_2,\cdots ,v_{n-k})$ is the supplementary set to 
$(u_1,u_2,\cdots ,u_k)$ in $(1,2,3,\cdots ,n)$,
and it satisfies 
$v_1<v_2<\cdots <v_{n-k}$, i.e. the relative order of 
these $n-k$ gluons in $(G_{v_1}^{\prime
}G_{v_2}^{\prime }\cdots G_{v_{n-k}}^{\prime })$ remains the same as in 
$(QG_1G_2\cdots G_n)$; $\sum_{\{u_1,\cdots,u_k\}}$ 
sums over all decompositions of $(12,\cdots ,n)$ 
into $(u_1,u_2,\cdots ,u_k)$ and $(v_1,v_2,\cdots ,v_{n-k})$. 
Note that $\left| Tr(QG_1^{\prime }G_2^{\prime }
\cdots G_n^{\prime})\right\rangle $ is just 
$\left| 1_{q1}1_{12}1_{23}\cdots 1_{i,i+1}\cdots
1_{n\overline{q}}\right\rangle $. Thus
$\left| Tr(QG_1G_2\cdots
G_n)\right\rangle $ can also be written as: 
\begin{equation}
\label{trq1}
\begin{array}{l}
\left| Tr(QG_1G_2\cdots G_n)\right\rangle =\left| 1_{q1}1_{12}1_{23}\cdots
1_{i,i+1}\cdots 1_{n\overline{q}}\right\rangle  \\ 
+\sum_{k=1}^n\sum_{\{u_1,\cdots,u_k\}}(-\frac
1{N_c})^k\left| 1_{u_1}1_{u_2}\cdots 1_{u_k}\right\rangle \left|
1_{qv_1}1_{v_1v_2}1_{v_2v_3}\cdots 1_{v_iv_{i+1}}\cdots 1_{v_{n-k}
\overline{q}}\right\rangle 
\end{array}
\end{equation}
We find immediately
$$
\langle 1_{q1}1_{12}1_{23}\cdots 1_{i,i+1}\cdots 1_{n\overline{q}}\mid
Tr(QG_1^{\prime }G_2^{\prime }\cdots G_n^{\prime })\rangle =N_c^{n+1} 
$$
Let us calculate
$$
\langle 1_{q1}1_{12}1_{23}\cdots 1_{i,i+1}\cdots 1_{n\overline{q}}\mid
1_u1_{q1}1_{12}1_{23}\cdots 1_{u-2,u-1}1_{u-1,u+1}\cdots 
1_{n\overline{q}}\rangle, 
$$
This is one of the terms in (\ref{trq1}) with $k=1$. Therefore we denote it 
$\langle f\mid k=1\;term\rangle $ in short, 
We draw a diagram as shown in
Fig.2 where corresponding to each $1_{st}$, there is a line starting from
position $s$ in the lower row to position $t$ in the upper row. Counting the
number of closed paths, we obtain
\begin{equation}
\label{in1q1-1}
\begin{array}{l}
\langle 1_{q1}1_{12}1_{23}\cdots 1_{i,i+1}\cdots 1_{n\overline{q}}
\mid 1_u1_{q1}1_{12}1_{23}\cdots 1_{u-2,u-1}1_{u-1,u+1}\cdots 1_{n
\overline{q}}\rangle \\
=N_c^{n+1-2+1}=N_c^n
\end{array}
\end{equation}

It is a little more complicated to calculate
\begin{equation}
\begin{array}{l}
\langle f\mid k=2\;term\rangle \\
=\langle 1_{q1}1_{12}1_{23}\cdots
1_{i,i+1}\cdots 1_{n\overline{q}}\mid
1_{u_1}1_{u_2}(1_{qv_1}1_{v_1v_2}1_{v_2v_3}\cdots 1_{v_iv_{i+1}}\cdots
1_{v_{n-k}\overline{q}})\rangle
\end{array}
\end{equation} 
There are two cases: one where $u_1$ and $u_2$ are adjacent to each other,
the other where
$u_1$ and $u_2$ are not adjacent. To see more clearly, we
look at the two cases separately, and we show 
the corresponding diagrams in Fig.3.
For the
case 
when $u_1$ and $u_2$ are adjacent, we immediately have
$$
\langle f\mid k=2\;term\rangle =N_c^{n+1-3+1}=N_c^{n-1} 
$$
When $u_1$ and $u_2$ are not neighbors, we find 
$$
\langle f\mid k=2\;term\rangle =N_c^{n+1-2\times 2+2}=N_c^{n-1} 
$$
Thus we obtain the same value in
both cases. 

As a matter of fact, for a general expression 
\begin{equation}
\begin{array}{l}
\langle f\mid k\;term\rangle \\
\equiv \langle 1_{q1}1_{12}1_{23}\cdots
1_{i,i+1}\cdots 1_{n\overline{q}}\mid (1_{u_1}1_{u_2}\cdots
1_{u_k})(1_{qv_1}1_{v_1v_2}1_{v_2v_3}\cdots 1_{v_iv_{i+1}}\cdots 1_{v_{n-k}
\overline{q}})\rangle \  
\end{array}
\end{equation}
where $k=3,4,5,\cdots ,n$, there is a unique value regardless of whether 
$u_1,u_2,\cdots ,u_k$ or part of them are neighbors.
Let us distinguish two cases with and without adjacent parton
labels. If no label in  $u_1,u_2,\cdots ,u_k$ is
adjacent to another, we have
$$
\langle f\mid k\;term\rangle =N_c^{n+1-2k+k}=N_c^{n-k+1} 
$$
where $n+1-2k$ is the number of double-line loops
and $k$ is the number
of closed paths involving $u_1,u_2,\cdots ,u_k$. 
If there are $m$ labels, 
each of which is adjacent to at least one other,
$m$ can be grouped into 
$l$ non-adjacent segments where labels 
belonging to the same segment are
continuous, i.e. $m=\sum_{i=1}^lm_i$ where 
$m_i$ is the number of labels in
the $i$-th segment. Hence we get the result 
$\langle f\mid k\;term\rangle =N_c^\epsilon $ 
where $\epsilon $ is
\begin{equation}
\begin{array}{l}
\epsilon =[n+1-2(k-m)-(m_1+1)-(m_2+1)-\cdots (m_l+1)]\\
+[k-m]+l=n-k+1 
\end{array}
\end{equation}
Here the first term is the number of double-line
loops, the second
term is the contribution from $k-m$ separated labels and the third term is
from $l$ continuous segments. Finally we have,
\begin{equation}
\label{1q1}
\begin{array}{l}
\langle 1_{q1}1_{12}1_{23}\cdots 1_{i,i+1}\cdots 1_{n
\overline{q}}\mid Tr(QG_1G_2\cdots G_n)\rangle  \\ 
=N_c^{n+1}+\sum_{k=1}^nC_n^k(-1)^k\frac 1{N_c^k}N_c^{n-k+1} \\ 
=N_c^{n+1}+\sum_{k=1}^nC_n^k(-1)^kN_c^{n-2k+1}
\end{array}
\end{equation}
where $C_n^k=\frac{n!}{k!(n-k)!}$ 
denotes the number of ways to pick $k$ out of $n$ labels.

Now we start calculating the general inner product $\langle f\mid
Tr(QG_{P(1)}G_{P(2)}\cdots G_{P(n)})\rangle $. First we focus on one of the
simplest cases: 
\begin{equation}
\label{ftr}
\langle f\mid Tr(QG_1G_3G_2G_4G_5\cdots G_n)\rangle 
\end{equation}
where two adjacent gluon labels 2 and 3 
are interchanged compared to (\ref{ftrq}). 
Similar to (\ref{trq}), we expand 
$\mid Tr(QG_1G_3G_2G_4G_5\cdots G_n)\rangle $ as follows: 
\begin{equation}
\label{trq2}
\begin{array}{l}
\left| Tr(QG_1G_3G_2G_4G_5\cdots G_n)\right\rangle  \\ 
=\left| Tr(QG_1^{\prime }G_3^{\prime }G_2^{\prime }G_4^{\prime }G_5^{\prime
}\cdots G_n^{\prime })\right\rangle \\
+\sum_{k=1}^n\sum_{\{u_1,\cdots,u_k\}}
(-\frac 1{N_c})^k\left|
1_{u_1}1_{u_2}\cdots 1_{u_k}\right\rangle \left| Tr(QG_{v_1}^{\prime
}G_{v_2}^{\prime }\cdots G_{v_{n-k}}^{\prime })\right\rangle 
\end{array}
\end{equation}
According to (\ref{inner2}), the first term of (\ref{ftr}) is: 
$$
\begin{array}{l}
\langle 1_{q1}1_{12}1_{23}\cdots 1_{i,i+1}\cdots 1_{n
\overline{q}}\mid 1_{q1}1_{13}1_{32}1_{24}1_{45}\cdots 1_{i,i+1}\cdots 1_{n
\overline{q}}\rangle  \\ 
=N_c^{n+1}/N_c^2=N_c^{n-1}
\end{array}
$$
Secondly we consider the
following $k=1$ terms: 
\begin{equation}
\label{inq1-2}(-1/N_c)\langle 1_{q1}1_{12}1_{23}\cdots 1_{i,i+1}\cdots 1_{n
\overline{q}}\mid 1_2(1_{q1}1_{13}1_{34}\cdots 1_{i,i+1}\cdots 
1_{n\overline{q}})\rangle 
\end{equation}
and 
\begin{equation}
\label{inq1-3}(-1/N_c)\langle 1_{q1}1_{12}1_{23}\cdots 1_{i,i+1}\cdots 1_{n
\overline{q}}\mid 1_3(1_{q1}1_{12}1_{24}\cdots 1_{i,i+1}\cdots 
1_{n\overline{q}})\rangle 
\end{equation}
where gluon 2 and 3 are picked out as singlet $1_2$ and $1_3$ respectively.
Note that the only difference between $\left| Tr(QG_1G_3G_2G_4G_5\cdots
G_n)\right\rangle $ and $\left| Tr(QG_1G_2G_3G_4G_5\cdots G_n)\right\rangle $
is that the order of gluon 2 and 3 is interchanged.
Thus
we also encounter (\ref{inq1-2}) and (\ref{inq1-3}) in 
calculating $\langle f\mid Tr(QG_1G_2\cdots G_n)\rangle $. From 
(\ref{in1q1-1}) we see that
(\ref{inq1-2}) and (\ref{inq1-3}) give the same value 
$-N_c^{n-1}$. For other terms we have
\begin{equation}
\label{inq1-u}
(-1/N_c)\langle f\mid 1_u(1_{q1}1_{13}1_{32}1_{24}1_{45}\cdots
1_{u-2,u-1}1_{u-1,u+1}\cdots 1_{n\overline{q}})\rangle =-N_c^{n-3}\ 
\end{equation}
where $u\neq 2,3$. We see that 
(\ref{inq1-u}) is suppressed by an additional factor
$1/N_c^2$ compared to (\ref{inq1-2}) and (\ref{inq1-3}). 
It is easy to show that the contribution from terms 
$$
(-\frac 1{N_c})^k\left| 1_{u_1}1_{u_2}\cdots 1_{u_k}\right\rangle \left|
Tr(QG_{v_1}^{\prime }G_{v_2}^{\prime }\cdots G_{v_{n-k}}^{\prime
})\right\rangle  
$$
with $k>1$ are suppressed at least by $1/N_c^2$ as
compared to (\ref{inq1-u}). Thus,
up to the highest order, we have:
\begin{equation}
\label{ftrq1}
\langle f\mid Tr(QG_1G_3G_2G_4G_5\cdots G_n)\rangle =
-N_c^{n-1}+O(N_c^{n-3})
\end{equation}

There are $(n-1)$ trace terms in 
$|H_c\rangle $, which differ from $|f\rangle $  
only in the order of two adjacent gluons. 
Their inner products with $|f\rangle $ are
given by (\ref{ftrq1}). 
There 
is a set of trace terms for which the labels of two 
non-adjacent gluons are interchanged compared to 
$| f\rangle $. We can in the same way prove 
that the projection of $| f\rangle $ on these 
trace terms gives $N_c^{n-1}+O(N_c^{n-3})$,  
which is in the same 
magnitude as (\ref{ftrq1}) 
but opposite in sign. 
These are next-to-leading order terms ($\sim N_c^{n-1}$) 
compared to the leading term of the order $N_c^{n+1}$ 
in (\ref{1q1}).
Remaining terms
have inner products which 
are suppressed by more powers of $1/N_c$. 

Corresponding to a chain state $|f\rangle $, 
we can classify all trace terms in
$|H_c\rangle $ into three groups. One is 
the leading term where the gluon order is the same as that in
$|f\rangle $. The second group are 
the next-to-leading terms,
where the order of two gluons is interchanged relative 
to $|f\rangle $. This group can be 
further classified into two subgroups according to 
whether the order of two adjacent or non-adjacent gluons 
is interchanged respectively. 
The third group contains higher order terms, 
in which the order of the gluons
differs even more from $| f\rangle $.  
The leading term is denoted as 
$L(f)$, the next-to-leading terms as $NL(f)$ with 
$NL_1(f)$ and $NL_2(f)$ for the adjacent and non-adjacent case 
respectively, and higher terms as $H(f)$. 
Hence, for the state
$$
\left| f\right\rangle =N_c^{-(n+1)/2}\left| 1_{q1}1_{12}1_{23}\cdots 
1_{i,i+1}\cdots 1_{n\overline{q}}\right\rangle  
$$
we find that, up to 
next-to-leading order, 
the projection on $|H_c\rangle $ is given by
\begin{equation}
\label{me1fhc}
\begin{array}{l}
\langle f\mid H_c\rangle 
= \sum_P(1/\sqrt{2})^nD^P\langle f\mid 
Tr(QG_{P(1)}G_{P(2)}\cdots G_{P(n)})\rangle  \\ 
\approx (1/\sqrt{2})^n\cdot N_c^{(n+1)/2}\cdot 
[(1-\frac n{N_c^2})D^{L(f)}
-\frac{1}{N_c^2}\sum _{P\in NL_1(f)}D^P+
\frac{1}{N_c^2}\sum _{P\in NL_2(f)}D^P ]
\end{array}
\end{equation}
We know from the previous
section that there are $n!$ singlet chain 
states which are denoted 
$\{|f_i\rangle , i=1,2,\cdots ,n!\}$. They are not orthogonal 
to each other. The largest inner product of two states is $1/N_c^2$.
Suppose we find a set of orthogonal states $\{|f'_i\rangle \}$ 
from $\{|f_i\rangle \}$. Up to 
next-to-leading order, we assume 
$\{|f'_i\rangle \}$ can be written in this form: 
\begin{equation}
\label{norm-st}
|f'_i\rangle \approx (1+\frac {C_1}{N_c^2})|f_i\rangle 
-\frac {C_2}{N_c^2}\sum _{j\in NL'(f_i)}|f_j\rangle 
\end{equation}
where $NL'(f_i)$ refers to the set of chain states which 
contribute to $|f'_i\rangle $ in 
next-to-leading order; $C_1$ and $C_2$ are 
constants of order 1. 
One can verify that 
the set $NL$ is included in $NL'$. The reason is that 
for a chain state the colors of the quark-antiquark 
pair play an equal role as those of gluons, while 
for a trace term in $|H_c\rangle$ it emerges as 
a color nonet which is different from the 
color octets of gluons. To see it more clearly, 
we take 
as an example
two permutations (01234) and (02341). 
Obviously for chain states 
$\langle 01234|02341\rangle =(1/N_c^2) 
\langle 01234|01234\rangle$, hence the orders  
(01234) and (02341) belong to 
the same $NL'$ set. 
Here we use a simplified notation 
for the chain state, e.g. 
$|01234\rangle \equiv 
|1_{01}1_{12}1_{23}1_{34}1_{40}\rangle $. 
It is easy to understand this because we can write the 
second state in 
the form $|10234\rangle $, which is different 
from the state $|01234\rangle $ in two labels 0 and 1. 
We can verify that
$\langle 01234|Tr(02341
)\rangle$ is 
suppressed by $O(1/N_c^4)$ relative to 
$\langle 01234|Tr(01234
)\rangle$, i.e. 
$|Tr(02341 
)\rangle$ does not belong to 
$NL(01234)$. 
Here we use a simplified notation 
for the chain state, e.g.  
$Tr(01234)\equiv Tr(QG_1G_2G_3G_4)$. 
Another different feature of $NL'$ 
compared to $NL$ is that the inner product 
of any two different states, where 
only two gluon labels are interchanged, is 
always suppressed by $1/N_c^2$ relative to the 
inner product of themselves,  
regardless 
of whether they belong to the adjacent or non-adjacent 
case. Hence $NL'$ can be written in this form: 
$NL'=NL_1+NL_2+\overline{NL}$. Up to $O(1/N_c^2)$ we have 
\begin{equation}
\begin{array}{rll}
\langle f'_i \mid H_c \rangle &=&
(1+\frac {C_1}{N_c^2})\langle f_i \mid H_c \rangle 
-\frac {C_2}{N_c^2}
\sum _{j\in NL(f_i)}\langle f_j \mid H_c \rangle \\
&=&2^{-n/2}N_c^{(n+1)/2}[(1+\frac {C_1}{N_c^2})
(1-\frac {n}{N_c^2})D^{L(f_i)}
-\frac{1}{N_c^2}\sum _{P\in NL_1(f_i)}D^P\\ 
&&+\frac{1}{N_c^2}\sum _{P\in NL_2(f_i)}D^P
-\frac{C_2}{N_c^2}\sum _{j\in NL(f_i)}D^{L(f_j)}]\\
&=&2^{-n/2}N_c^{(n+1)/2}[(1+\frac {C_1-n}{N_c^2})D^{L(f_i)}
-\frac{C_2+1}{N_c^2}\sum _{P\in NL_1(f_i)}D^P\\
&&-\frac{C_2-1}{N_c^2}\sum _{P\in NL_2(f_i)}D^P
-\frac{C_2}{N_c^2}\sum _{P\in \overline{NL}(f_i)}D^P\; ]
\end{array}
\end{equation}
and the projection square has the following form:
\begin{equation}
\begin{array}{l}
|\langle f'_i \mid H_c \rangle |^2
=2^{-n}N_c^{n+1}[\; (1+\frac {2(C_1-n)}{N_c^2})|D^{L(f_i)}|^2 
-\frac {2(C_2+1)}{N_c^2}
\sum _{P\in NL_1(f_i)}Re(D^{L(f_i)}\cdot D^{P*})\\
-\frac {2(C_2-1)}{N_c^2}
\sum _{P\in NL_2(f_i)}Re(D^{L(f_i)}\cdot D^{P*})
-\frac{2C_2}{N_c^2}
\sum _{P\in \overline{NL}(f_i)}Re(D^{L(f_i)}\cdot D^{P*})\; ]
\end{array}
\end{equation}
where we only keep the next-to-leading order. 
The total sum is 
\begin{equation}
\label{me2}
\begin{array}{l}
\sum _{i=1}^{n!}|\langle f'_i \mid H_c \rangle |^2 \\ 
=2^{-n}N_c^{n+1}
[\; (1+\frac {2(C_1-n)}{N_c^2})\sum _P|D^P|^2-\frac{4(C_2+1)}{N_c^2}
\sum _{\{P,P'\}\in NL_1}Re(D^P\cdot D^{P'*})\\
-\frac{4(C_2-1)}{N_c^2}
\sum _{\{P,P'\}\in NL_2}Re(D^P\cdot D^{P'*})
-\frac{4C_2}{N_c^2}
\sum _{P\in \overline{NL}(f_i)}Re(D^{L(f_i)}\cdot D^{P*})\;]
\end{array}
\end{equation}
where $\{P,P'\}\in NL_1$ means that the orders
$P$ and $P'$ are different 
in only two neighboring
gluon labels, $\{P,P'\}\in NL_2$ means that 
they are different in two non-adjacent gluon labels, 
and $\{P,P'\}\in \overline{NL}$ means $\{P,P'\}\in NL'$ 
but not included in $NL_{1}+NL_{2}$.

Now we start calculating 
$\sigma _{tree}(e^{+}e^{-}\rightarrow q\overline{q}+ng)$ 
in the large $N_c$ limit. Recall
that the ordinary matrix element is given by
$$
M_{ab}^{\alpha _1\alpha _2\cdots \alpha _n}=
\sum_P(T^{\alpha _{P(1)}}
T^{\alpha _{P(2)}}\cdots T^{\alpha _{P(n)}})_{ab}D^P 
$$
The total cross section is then: 
\begin{equation}
\label{sig-tree}
\begin{array}{l}
\sigma _{tree}(e^{+}e^{-}\rightarrow q
\overline{q}+ng) \\ =\sum_{a,b,\alpha _1,\alpha _2,\cdots ,\alpha
_n}\int d\Omega \left| M_{ab}^{\alpha _1\alpha _2\cdots
\alpha _n}\right| ^2 \\ 
=\sum_{a,b,\alpha _1,\alpha _2,\cdots ,\alpha
_n}\int d\Omega \left| \sum_P(T^{\alpha _{P(1)}}T^{\alpha
_{P(2)}}\cdots T^{\alpha _{P(n)}})_{ab}D^P\right| ^2
\end{array}
\end{equation}
In evaluating (\ref{sig-tree}), we mainly encounter two types of
traces of Gell-Mann matrices
\begin{equation}
\label{trta1}
\begin{array}{l}
Tr(T^{\alpha _1}T^{\alpha _2}\cdots T^{\alpha _n}T^{\alpha _n}\cdots
T^{\alpha _2}T^{\alpha _1}) \\ 
=N_c\cdot C_F^n=N_c(\frac{N_c^2-1}{2N_c})^n \\ 
\approx (1/2)^n\cdot N_c^{(n+1)}(1-\frac n{N_c^2}) 
\end{array}
\end{equation}
and 
\begin{equation}
\label{trat}
\begin{array}{rll}
Tr(A_1T^\alpha A_2T^\alpha A_3)
&=&\frac 12Tr(A_2)Tr(A_1A_3)-\frac 1{2N_c}Tr(A_1A_2A_3)\\
Tr(A_1T^{\alpha})Tr(A_2T^{\alpha})&=&\frac 12Tr(A_1A_2)
-\frac 1{2N_c}Tr(A_1)Tr(A_2) 
\end{array}
\end{equation}
where $A_1$,$A_2$ and $A_3$ are chains 
of products of Gell-Mann matrices.
The result in (\ref{trta1}) is just the color factor 
associated with the terms $\left| D^P\right|^2$,  
while (\ref{trat})
gives the color factor for the kinematic interference terms
$D^P\cdot D^{P^{\prime }}$, 
where $P$ and 
$P^{\prime }$ denote two
different permutations. In the large $N_c$ limit the
factor (\ref{trta1}) is the 
leading one. The next-to-leading contribution 
comes from $D^P\cdot D^{P'}$ terms 
where $P$ and $P'$ are different in only two 
gluon labels. 
Also the next-to-leading contributions can be 
classified into two different cases, 
one is that $\{ P,P'\}\in NL_1$, i.e. 
$P$ and $P'$ are different in two 
neighboring gluon labels, the other 
is that $\{ P,P'\}\in NL_2$, i.e.  
they are different in two non-adjacent 
gluon labels. For $\{ P,P'\}\in NL_1$,  
the color factor is
$-\frac 1{2N_c}\cdot N_c\cdot C_F^{n-1}\approx 
-\frac 1{2^{n}}N_c^{n-1}$,  
while for $\{ P,P'\}\in NL_2$, the color factor is 
$\frac 1{2^{n}}N_c^{n-1}$ in the large $N_c$ limit.  
After keeping terms up to 
next-to-leading order we have
\begin{equation}
\label{sig-tree1}
\begin{array}{l}
\sigma _{tree}(e^{+}e^{-}\rightarrow q
\overline{q}+ng) \\ 
=(1/2)^n\cdot N_c^{(n+1)}\cdot
\int d\Omega [\;(1-\frac n{N_c^2})\sum_P|D^P| ^2-\frac{2}{N_c^2}
\sum _{\{P,P'\}\in NL_1}Re(D^P\cdot D^{P'*})\\
+\frac{2}{N_c^2}
\sum _{\{P,P'\}\in NL_2}Re(D^P\cdot D^{P'*})\; ]
\end{array}
\end{equation}

We write the fraction of singlet chain states as: 
\begin{equation}
\label{p-sss1}
\begin{array}{l}
P(e^{+}e^{-}\rightarrow q 
\overline{q}+ng\rightarrow SCS) \\ 
=(\int d\Omega 
\sum _{i=1}^{n!}|\langle f'_i \mid H_c \rangle |^2 )/\sigma
_{tree}(e^{+}e^{-}\rightarrow q\overline{q}+ng) \end{array}
\end{equation}
where $\sum _{i=1}^{n!}|\langle f'_i \mid H_c \rangle |^2 $ 
is given by (\ref{me2}). 
According to Eq.(\ref{me2}, \ref{sig-tree1}, \ref{p-sss1}) and 
we reach our final result up to next-to-leading order
\begin{equation}
\label{result}
\begin{array}{l}
P(e^{+}e^{-}\rightarrow q\overline{q}+ng\rightarrow SCS)\\
=1+\frac {1}{N_c^2}[(2C_1-n)-(4C_2+2)T_1-(4C_2-2)T_2-4C_2T_3]\\
\end{array}
\end{equation}
where $T_1$ and $T_2$ are defined by:
\begin{eqnarray}
T_1=\frac{\int d\Omega \sum _{\{P,P'\}\in NL_1}Re(D^P\cdot D^{P'*})}
{\int d\Omega \sum_P\left| D^P\right| ^2}\nonumber\\
T_2=\frac{\int d\Omega \sum _{\{P,P'\}\in NL_2}Re(D^P\cdot D^{P'*})}
{\int d\Omega \sum_P\left| D^P\right| ^2}\nonumber \\
T_3=\frac{\int d\Omega 
\sum _{\{P,P'\}\in \overline{NL}}Re(D^P\cdot D^{P'*})}
{\int d\Omega \sum_P\left| D^P\right| ^2}
\end{eqnarray}
The result in Eq.(\ref{result}) gives the fraction of 
events, for which the colors of the gluons 
correspond to a single chain from the quark 
to the antiquark. We expect that these states 
hadronize producing a corresponding chain of 
hadrons. The remaining events correspond to 
more complicated color structures, where one 
gluon can be connected to more than two other 
gluons, as indicated in Fig. 2 and 3. Some 
dynamical feature of the confining mechanism may 
imply that also these states result in string-like 
hadronic final states, but it is also conceivable 
that these parton states can produce more 
complex hadron configurations.

\section{$H_c$ in soft gluon bremsstrahlung and 
its relation to dipole cascade model}
\label{soft-gluon}

We have not yet touched the momentum function $D$ in $H_c$ so far. 
The number and complexity of Feynman diagrams increases drastically 
with growing number of emitted gluons in normal situations. However, 
in the case of soft gluon bremsstrahlung, 
the $D$ function has a simple 
and recursive structure. The method we use to derive the $D$ function or 
$H_c$ in this section is called soft gluon insertion 
technique\cite{BCM}. By recursively adding a softer (with lower energy) 
gluon in multigluon emissions, the distribution for each new 
emission is approximately determined by an eikonal current 
stemming from all the harder gluons. The result factorizes 
between the emissions and is equivalent to classical 
bremsstrahlung under certain angular ordering conditions.

Assume that in 
$e^+e^-$ annihilation, 
$e^+e^-\rightarrow 
q\overline{q}g_1g_2\cdots g_n$ 
, $n$ gluons are all soft ones and 
their energies/momenta are strongly ordered:
\begin{equation}
\label{e-order}
E_p\sim E_{p'}\gg E_{k_1}\gg E_{k_2}\cdots \gg E_{k_n}
\end{equation}
where $p,p',k_1,k_2,\cdots ,k_n$ are 4-momenta of 
$q,\overline{q},g_1,g_2,\cdots ,g_n$ respectively.
When the hardest gluon $g_1$ is emitted, it has two legs to attach to, 
one is the quark's momentum  
and the other is the anti-quark's. They give rise to the amplitude:
\begin{equation}
M_{ab}^{\alpha _1}
\sim g_s\varepsilon _{\mu _1}J_{ab\alpha _1}^{\mu _1}
=g_s(\frac{p^{\mu _1}}{p\cdot k_1}-
\frac{p'^{\mu _1}}{p'\cdot k_1})T_{ab}^{\alpha _1}
\varepsilon _{\mu _1}
\equiv g_sJ^{\mu _1}(k_1;p,p')T_{ab}^{\alpha _1}
\varepsilon _{\mu _1}
\end{equation}
where $\varepsilon$ denotes the gluon's polarization 4-vector 
and $g_s$ is the strong coupling constant.

When the second hardest gluon $g_2$ is emitted, it has three legs to 
attach to: $q$, $\overline{q}$ and $g_1$. The amplitude is:
\begin{equation}
\label{j1}
\begin{array}{l}
M_{ab}^{\alpha _1\alpha _2}
\sim g_s^2\varepsilon _{\mu _1}\varepsilon _{\mu _2}
J_{ab\alpha _1\alpha _2}^{\mu _1\mu _2}\\
=g_s^2\varepsilon _{\mu _1}\varepsilon _{\mu _2}
J^{\mu _1}(k_1;p,p')[\frac{p^{\mu _2}}{p\cdot k_2}
(T^{\alpha _2}T^{\alpha _1})_{ab}-\frac{p'^{\mu _2}}{p'\cdot k_2}
(T^{\alpha _1}T^{\alpha _2})_{ab}+\frac{k_1^{\mu _2}}{k_1\cdot k_2}
T^{\alpha _2}_{A\alpha _1\beta}T^{\beta}_{ab}]\\
=g_s^2\varepsilon _{\mu _1}\varepsilon _{\mu _2}
[J^{\mu _1}(k_1;p,p')J^{\mu _2}(k_2;p,k_1)
(T^{\alpha _2}T^{\alpha _1})_{ab}\\
+J^{\mu _1}(k_1;p,p')J^{\mu _2}(k_2;k_1,p')(T^{\alpha _1}T^{\alpha _2})_{ab}]
\end{array}
\end{equation}
where $T^{\alpha _2}_{A\alpha _1\beta}=i f_{\alpha _1\alpha _2\beta}$ is 
the generator of the adjoint representation of $SU(3)$, and the rule 
$i f_{\alpha _1\alpha _2\beta}T^{\beta}=
T^{\alpha _1}T^{\alpha _2}-T^{\alpha _2}T^{\alpha _1}$ has been used. 

In the case of $n$-gluon emission, we can prove in 
the same way that the amplitude and its corresponding 
$|H_c\rangle $ can be written in this form:
\begin{equation}
\label{jn}
\begin{array}{rll}
M_{ab}^{\alpha _1\alpha _2\cdots \alpha _n}
&=&\sum_P(T^{\alpha _{P(1)}}T^{\alpha _{P(2)}}
\cdots T^{\alpha _{P(n)}})_{ab}D^P \\
&\sim &g_s^n\varepsilon _{\mu _1}
\varepsilon _{\mu _2}\cdots \varepsilon _{\mu _n}
[\sum_P(T^{\alpha _{P(1)}}T^{\alpha _{P(2)}}
\cdots T^{\alpha _{P(n)}})_{ab}\\
&&\cdot J^{\mu _1}(k_1;p,p')J^{\mu _2}(k_2;k_{2h},k_{2e})\cdots 
J^{\mu _n}(k_n;k_{nh},k_{ne})]\\
|H_c\rangle &\sim &g_s^n 
\varepsilon _{\mu _1}\varepsilon _{\mu _2}\cdots \varepsilon _{\mu _n}
\sum_P(1/\sqrt{2})^n|Tr(QG_{P(1)}
G_{P(2)}\cdots G_{P(n)})\rangle \\
&&\cdot J^{\mu _1}(k_1;p,p')J^{\mu _2}(k_2;k_{2h},k_{2e})\cdots 
J^{\mu _n}(k_n;k_{nh},k_{ne})
\end{array}
\end{equation}
where subscripts $ih$ and $ie$ (where $i=2,\cdots ,n$)
are determined by the following procedure:
in the sequence $(0,P(1),P(2),\cdots ,P(n),0)$ 
(where we imply $0\equiv q$ at the head and $0\equiv \overline{q}$
at the end), find the position of $i$, 
take away all greater numbers in the sequence, 
the left-nearest neighbor 
to $i$ is $ih$ and its right-nearest neighbor 
is $ie$. Having this complete form 
of $|H_c\rangle $, we can calculate 
the fraction of any color state by projecting the state onto it. 
Here we are only interested in what happens to SCS 
in the large $N_c$ limit. We consider a chain state 
$|f_P\rangle$ which 
corresponds to a specific order of gluons: $(P(1),P(2),\cdots ,P(n))$ 
where $P$ denotes the permutation of $(1,2,\cdots ,n)$. 
According to the former section, in the leading order, 
the inner product $\langle f_P\mid H_c\rangle$ only picks up the term 
with the same order of gluons in $|H_c\rangle $:
\begin{equation}
\begin{array}{rll}
\langle f_P\mid H_c\rangle 
&\simeq &
g_s^n\varepsilon _{\mu _1}\varepsilon _{\mu _2}
\cdots \varepsilon _{\mu _n}
J^{\mu _1}(k_1;p,p')J^{\mu _2}(k_2;k_{2h},k_{2e})
\cdots J^{\mu _n}(k_n;k_{nh},k_{ne})\\
&&\cdot (1/\sqrt{2})^n\langle f_P\mid 
Tr(QG_{P(1)}G_{P(2)}\cdots G_{P(n)})\rangle
\end{array}
\end{equation}
According to (\ref{me1fhc}), the projection square is: 
\begin{equation}
\label{fPhc2}
\mid \langle f_P\mid H_c\rangle \mid ^2=N_c^{n+1}g_s^{2n}
(p,p')_{k_1}(k_{2h},k_{2e})_{k_2}\cdots (k_{nh},k_{ne})_{k_n}
\end{equation} 
where the 
antenna term is defined by:
\begin{equation}
(p_1,p_2)_{k}\equiv \frac {p_1\cdot p_2}{(k\cdot p_1)(k\cdot p_2)}
\end{equation}
Hence we see that each gluon $g_i$ is
associated with two harder gluons $g_{ih}$ and $g_{ie}$ which 
are nearest to its position in the sequence.

As an example, we look at the case:
\begin{equation}
\left| f\right\rangle =N_c^{-(n+1)/2}\left| 1_{q1}1_{12}1_{23}\cdots 
1_{ii+1}\cdots 1_{n\overline{q}}\right\rangle 
\end{equation} 
with gluons' order $(1,2,\cdots ,n)$. The projection square is:
\begin{equation}
\mid \langle f\mid H_c\rangle \mid ^2=N_c^{n+1}g_s^{2n}
(p,p')_{k_1}(k_1,p')_{k_2}(k_2,p')_{k_3}\cdots (k_{n-1},p')_{k_n}
\end{equation}
The cross section of 
$e^{+}e^{-}\rightarrow q\overline{q}
+ng\rightarrow \left| f\right\rangle$ is:
\begin{equation}
\begin{array}{l}
d\sigma (e^{+}e^{-}\rightarrow \left| f\right\rangle)=N_c^{n+1}g_s^{2n}
\frac{d^3p}{(2\pi )^32E}\frac{d^3p'}{(2\pi )^32E'}
\prod _{i=1}^n\frac{d^3k_i}{(2\pi )^32E_{i}}\\
(p,p')_{k_1}(k_1,p')_{k_2}(k_2,p')_{k_3}\cdots (k_{n-1},p')_{k_n}
(2\pi )^4\delta ^{(4)}(P_{e^+}+P_{e^-}-[p+p'+\sum k_i])
\end{array}
\end{equation} 
In the approximation of soft gluon bremsstrahlung, 
the above cross section can be written as:
\begin{equation}
\label{d-sigma}
\begin{array}{l}
d\sigma (e^{+}e^{-}\rightarrow \left| f\right\rangle)
=d\sigma (e^{+}e^{-}\rightarrow q\overline{q})
\prod _{i=1}^n \frac{N_c}{2}\alpha _s\frac{d^3k_i}{4\pi ^24E_{i}}
\frac {k_{i-1}\cdot p'}{(k_i\cdot k_{i-1})(k_i\cdot p')}\\
=d\sigma (e^{+}e^{-}\rightarrow q\overline{q})
\prod _{i=1}^n \frac{N_c}{2\pi}\alpha _s
\frac{d(k_i\cdot k_{i-1})}{(k_i\cdot k_{i-1})}
\frac{d(k_i\cdot p')}{(k_i\cdot p')}
\end{array}
\end{equation} 
where the strong coupling constant 
$\alpha _s\equiv \frac{g_s^2}{4\pi}$, and 
we use the notation $k_0\equiv p$ for convenience. 
Define two new variables for the gluon $g_i$ 
which are called the generalized 
rapidity and the transverse momentum: 
\begin{equation}
\begin{array}{l}
y(g_{i-1}\overline{q}\rightarrow g_{i})
=\frac 12ln(\frac{k_i\cdot k_{i-1}}{k_i\cdot p'})
\;,\;\;
k^2_{T}(g_{i-1}\overline{q}\rightarrow g_{i})
=\frac{(k_i\cdot k_{i-1})(k_i\cdot p')}{k_{i-1}\cdot p'}
\end{array}
\end{equation}
We can rewrite Eq. (\ref{d-sigma}) in terms of 
$y$ and $p_{T}$:
\begin{equation}
\label{dipole}
d\sigma (e^{+}e^{-}\rightarrow \left| f\right\rangle)
=d\sigma (e^{+}e^{-}\rightarrow q\overline{q})
\prod _{i=1}^n d\sigma (g_{i-1}\overline{q}\rightarrow g_{i})
\end{equation} 
where we define the emission rate for $g_i$:
\begin{equation}
\label{dipole2}
d\sigma (g_{i-1}\overline{q}\rightarrow g_{i})
=\frac{N_c}{2\pi}\alpha _s 
dy(g_{i-1}\overline{q}\rightarrow g_{i})
\frac{dk_{T}^2(g_{i-1}\overline{q}\rightarrow g_{i})}
{k_{T}^2(g_{i-1}\overline{q}\rightarrow g_{i})}
\end{equation} 
Eq. (\ref{dipole}) 
is the result for one chain state with a specific order. 
If we take into account all $n!$ states, Eq.(\ref{dipole})
becomes:
\begin{equation}
\label{dipole1}
\begin{array}{rll}
d\sigma (e^{+}e^{-}\rightarrow \sum _{j=1}^{n!}
\left| f_j\right\rangle)
&=&d\sigma (e^{+}e^{-}\rightarrow q\overline{q})
d\sigma (q\overline{q}\rightarrow g_{1})\\
&&\cdot [d\sigma (qg_1\rightarrow g_2)
d\sigma (qg_2g_1\overline{q}\rightarrow g_{3}\cdots g_n)\\
&&+d\sigma (g_1\overline{q}\rightarrow g_2)
d\sigma (qg_1g_2\overline{q}\rightarrow g_{3}\cdots g_n)] 
\end{array}
\end{equation} 
Eq.(\ref{dipole2}, \ref{dipole1})
is just the dipole radiation formula in the 
Lund Dipole Cascade Model\cite{gustafson}.
In a more general situation the natural 
ordering variable is transverse momentum, 
and it is also possible to go beyond the 
eikonal approximationso that the Altarelli-Parisi 
splitting functions are properly reproduced 
for collinear emissions. In ref. \cite{gustafson} 
it is also demonstrated that the dipole 
formulation reproduces the angular 
ordering constraint due to soft gluon interference
\cite{angularordering}. This restricts 
interference effects and implies that 
gluon emission is a local process, 
in which only a limited set of gluons 
contribute to the emisson of a softer 
gluon (where soft should mean in the 
rest frame of the parent set).

We have derived the Lund dipole cascade 
model from the $H_c$ approach in the 
leading order. Now we begin to 
discuss the next-to-leading order. 
The general case in this order 
is rather complicated, and we therefore 
first discuss the simple cases with 
two and three gluons.  The local character 
of the dipole cascade emission implies, 
however, that these results may be 
relevant also for a more general situation. 

For $n=2$, i.e. 
when there are two gluons in the 
final state, we have two singlet chain states 
$|f_{1,2}\rangle =\{|012\rangle ,|021\rangle\}$ 
(the normalization factor is $N_c^{-3/2}$). 
We have $NL'=\{(012),(021)\}$, 
$NL_1=NL=NL'$ and $NL_2=\overline{NL}=\{\}$. 
There are two types of 
inner products between a chain 
state and a trace state in 
$|H_c\rangle $: 
$\langle 012|Tr(012)\rangle 
=N_c^3-2N_c+\frac 1{N_c}\approx N_c^3$
and $\langle 012|Tr(021)\rangle 
=-N_c+\frac 1{N_c}\approx -N_c$. 
Note that we have used a simplified notation
for the chain state and trace state, e.g.
$|01234\rangle \equiv 
|1_{01}1_{12}1_{23}1_{34}1_{40}\rangle $ and 
$Tr(01234)\equiv Tr(QG_1G_2G_3G_4)$ etc.. 
If we use orthogonalized chain states as shown in 
Eq.(\ref{norm-st}), we have:
\begin{equation}
\begin{array}{rll}
|(012)'\rangle 
&=&(1+\frac{C_1}{N_c^2})|012\rangle 
-\frac{C_2}{N_c^2}|021\rangle \\
|(021)'\rangle  
&=&(1+\frac{C_1}{N_c^2})|021\rangle
-\frac{C_2}{N_c^2}|012\rangle \\
\end{array}
\end{equation}
From Eq.(\ref{me2}) and taking into account 
the normalization factor for chain states, 
we obtain:
\begin{equation}
\begin{array}{l}
|\langle (012)'| H_c \rangle |^2 
+|\langle (021)'| H_c \rangle |^2 \\ 
=\frac 14N_c^{3}
[\; (1+2\frac {C_1}{N_c^2})(|D^{12}|^2+|D^{21}|^2)
-\frac{4(C_2+1)}{N_c^2} Re(D^{12}\cdot D^{21*})\; ]\\  
=N_c^{3}g_s^4\{\; (1+2\frac {C_1}{N_c^2})
(0,\overline{0})_1[\; (1,\overline{0})_2+(0,1)_2\;]\\
-\frac{4(C_2+1)}{N_c^2}(0,\overline{0})_1
[\; (0,\overline{0})_2
-(0,1)_2-(1,\overline{0})_2\;]\;\}
\end{array} 
\end{equation} 
where we have used a simplified notation for the 
antenna term, for example, $(0,1)_2\equiv (p,k_1)_{k_2}$ and 
$(1,\overline{0})_2\equiv (k_1,p')_{k_2}$ etc.. 
We see that the magnitude of $(1,\overline{0})_2$ and 
$(0,1)_2$ in leading order obtains a correction term which is 
of order $1/N_c^2$. There is also 
a 
negative term $(0,\overline{0})_2$ from the interference which 
is absent in leading order. This term corresponds to 
the emission rate of gluon 2 from the 
dipole $q\overline{q}$.

For $n=3$, there are six chain states:
\begin{equation} 
\{|f_i\rangle ,i=1\cdots 6\}
=\{(0123),(0132),(0213),(0312),(0231),(0321)\}
\end{equation}
Since $\langle f_i|f_j\rangle =1/N_c^2$ for any 
$i\neq j$, all these states belong to the same $NL'$. 
$NL_1$ and $NL_2$ are given by: 
\begin{equation}
\label{nl12}
\begin{array}{rll}
NL_{1,2}(0123)&=&\{(0213),(0132)\},\{(0321)\}\\
NL_{1,2}(0132)&=&\{(0312),(0123)\},\{(0231)\}\\
NL_{1,2}(0213)&=&\{(0123),(0231)\},\{(0312)\}\\
NL_{1,2}(0312)&=&\{(0132),(0321)\},\{(0213)\}\\
NL_{1,2}(0231)&=&\{(0321),(0213)\},\{(0132)\}\\
NL_{1,2}(0321)&=&\{(0231),(0312)\},\{(0123)\}
\end{array}
\end{equation}
and $\overline{NL}$ is obtained by $NL'-NL_1-NL_2$. 
We can write the above equation in another form. 
According to Eq.(\ref{nl12}), 
the following pairs of orders for chain states 
belong to $NL_1$, $NL_2$ and $\overline{NL}$, respectively: 
\begin{equation}
\label{nl1}
\begin{array}{rll}
NL_1&:\;\;& 
\{(0123),(0213)\},\;\;\{(0123),(0132)\},\;\;\{(0132),(0312)\}\\
&&\{(0213),(0231)\},\;\;\{(0312),(0321)\},\;\;\{(0231),(0321)\}
\end{array}
\end{equation}
\begin{equation}
\label{nl2}
NL_2: \;\; 
\{(0123),(0321)\},\;\;\{(0132),(0231)\},\;\;\{(0213),(0312)\}
\end{equation}
\begin{equation}
\label{nl3}
\begin{array}{rll}
\overline{NL}&:\;\;& 
\{(0123),(0312)\},\;\;\{(0123),(0231)\},\;\;\{(0312),(0231)\}\\
&&\{(0321),(0132)\},\;\;\{(0321),(0213)\},\;\;\{(0132),(0213)\}
\end{array}
\end{equation}
According Eq.(\ref{nl1}-\ref{nl3}), 
the leading contribution involves
\begin{equation}
\label{lead}
\begin{array}{rll}
\sum _{all\;P}|D^P|^2&=&2^3g_s^6
(0,\overline{0})_1\;\{\;(1,\overline{0})_2
\;[\;(0,1)_3+(1,2)_3+(2,\overline{0})_3\;]\\
&&+(0,1)_2\; [\;(0,2)_3+(2,1)_3+(1,\overline{0})_3\;]\;\}
\end{array}
\end{equation}
The next-to-leading contribution
consists of 
three parts which are from $NL_1$, $NL_2$ 
and $\overline{NL}$ respectively. The $NL_1$ 
contribution corresponds to 
the following sums:
\begin{equation}
\label{nextl1}
\begin{array}{l}
\sum _{P\in NL_1}Re(D^PD^{P'*})=2\;g_s^6\; 
(0,\overline{0})_1
\;\{\; -I_1(0,1,\overline{0}|2)I_1(1,\overline{0},2|3)\\
-I_1(0,1,\overline{0}|2)I_1(1,0,2|3) 
+2(1,\overline{0})_2\;I_1(1,2,\overline{0}|3) 
+2(1,\overline{0})_2\;I_1(0,1,2|3)\\
+2(0,1)_2\;I_1(2,1,\overline{0}|3)
+2(0,1)_2\;I_1(0,2,1|3)\;\}
\end{array}
\end{equation}
where we the interference pattern $I_1$ is defined by: 
\begin{equation}
\begin{array}{rll}
I_1(g_1,g_2,g_3|g_4)&\equiv& 
(g_1,g_3)_{g_4}-(g_1,g_2)_{g_4}-(g_2,g_3)_{g_4}\\
I_1(g_3,g_2,g_1|g_4)&=&I_1(g_1,g_2,g_3|g_4) 
\end{array}
\end{equation}
where $g_i$ is the label for gluon $i$. 
The $NL_2$ contribution corresponds to 
\begin{equation}
\label{nextl2}
\begin{array}{l}
\sum _{P\in NL_2}Re(D^PD^{P'*})=2\;g_s^6\; 
(0,\overline{0})_1 \;I_1(0,1,\overline{0}|2)\;\\
\cdot [\; I_1(0,2,\overline{0}|3)
+I_1(0,1,\overline{0}|3) 
-2\;(1,2)_3\;]
\end{array}
\end{equation}
The $\overline{NL}$ contribution corresponds to 
\begin{equation}
\label{nextl3}
\begin{array}{l}
\sum _{P\in \overline{NL}}Re(D^PD^{P'*})\\
=2\;g_s^6\; (0,\overline{0})_1
\;\{\; 2\;(1,\overline{0})_2\;I_2(0,1,2,\overline{0}|3)
+2\;(0,1)_2\;I_2(0,2,1,\overline{0}|3)\\
-I_1(0,1,\overline{0}|2)[\;I_1(1,2,\overline{0}|3)
+I_1(0,1,2|3)+I_1(0,2,1|3)+I_1(2,1,\overline{0}|3)\;]\;\} 
\end{array}
\end{equation}
where the interference pattern $I_2$ is defined by:
\begin{equation}
I_2(g_1,g_2,g_3,g_4|g_5)\equiv
(g_1,g_4)_{g_5}+(g_2,g_3)_{g_5}
-(g_1,g_3)_{g_5}-(g_2,g_4)_{g_5}
\end{equation}
Substituting the leading and next-to-leading 
contribution given in Eq.(\ref{lead}-\ref{nextl3}) 
into Eq.(\ref{me2}), we can 
obtain the cross section
for chain states up to 
 next-to-leading order.

In a strongly ordered gluon cascade the emission amplitude has
an eikonal form. 
In this case, we have shown that in leading 
order in $N_c$, the cross section of 
$e^{+}e^{-}\rightarrow q\overline{q}+ng$ can be decomposed 
into $n!$ independent or incoherent parts and each part 
represents the contribution from the chain state 
with a specific order of gluons.
The cross section of 
a singlet chain state exactly corresponds to 
the emission of gluons by a specific sequence of dipoles. 
The sequence is determined by the gluon order of 
the chain state. The softest gluon is emitted 
independently and in an equal emission probability 
by all possible dipoles each of 
which is stretched by two adjacent harder gluons,   
and each dipole corresponds to a specific chain state. 
Hence to sum over all possible chain states 
is the same thing as to sum over all possible dipole 
sequences. In next-to-leading order, 
the emission probability aquires a $O(1/N_c^2)$ 
correction and may not be necessarily equal for 
different neighbor-gluon dipoles. Furthermore there 
appear dipoles stretched by non-adjacent 
gluons which are separated by only one 
extra gluon.  
The next-to-leading corrections arise from 
interferences of two $D$-functions with 
their gluon orders in the $NL_{1,2}$ or $\overline{NL}$ 
set. For two $D$-functions belonging to $NL_{1,2}$,
only one interference pattern $I_1$ is relevant. The 
other interference pattern $I_2$ is associated with 
$D$-functions belonging to $\overline{NL}$. 
$I_1$ is related to three consecutive gluons and 
$I_2$ to four gluons which are not 
necessarily consecutive. 
In summary, each next-to-leading correction brought by 
the interference of a pair of 
$D$-functions, say, $D^P$ and $D^{P'}$, 
can be regarded as a small perturbation to the dipole sequence 
which corresponds to $|D^P|^2$ or 
$|D^{P'}|^2$ by having dipoles formed by non-adjacent 
gluons while keeping the rest of the dipole 
sequence same as that corresponding to $|D^P|^2$ or 
$|D^{P'}|^2$. Compared to higher-order corrections, 
the next-to-leading correction causes the least 
perturbation to the dipole sequence of the leading order.

\section{Summary and Conclusion}

The phenomenological color flow picture,
commonly used in the Lund model and
the cluster model,
is to assign the color connection of a final parton
system. In these models, for an $e^{+}e^{-}\rightarrow q\overline{q}+ng$
event, the neutral color flow is definitely determined and begins at the
quark, connects each gluon one by one in a certain order, and ends at the
antiquark. Each flow piece spanned between two partons is color-neutral and
its hadronization is treated in 
a way similar to a $q\overline{q}$ singlet
system. The present hadronization models work successfully, which
shows that this
picture is a good approximation to the real 
world. In this paper we use the 
method of color Hamiltonian, a strict 
formulation developed from PQCD, to
study the structure of chain states in 
$e^{+}e^{-}\rightarrow q\overline{q}+ng$ for
finite $N_c=3$ and in
the large $N_c$ limit. 
For large $N_c$ these states correspond to well-defined color
topologies. They just correspond to
the phenomenological neutral color flow. 
Therefore we may expect that the fraction of 
the non-chain state is an estimate 
of the fraction of events,
where color reconnection 
is possible. 
It is also conceivable that color structures, 
where one gluon is connected to more than 
two other gluons, may result in more complicated 
hadron configurations. (The problem of finding 
experimental observables for these final states is, 
however, beyond the scope of this paper.)

There are $n!$ chain states,
each of which corresponds to a specific
order of $n$ gluons. 
When $N_c$ is 3, as in the real 
world, any two different chain states are not orthogonal to
each other. To derive the total fraction for chain states
we must orthogonalize them. We give two types 
of recipes of orthogonalization: 
one is to symmetrize the original chain states, the other is 
to find the transformation matrix which 
slightly differs from the unit matrix,
by exploiting the fact that every
two different chain states are approximately 
orthogonal up to order $\frac 1{N_c^2}$. 
As an example, we give the numerical result 
for the rate of chain states
in $e^{+}e^{-}\rightarrow q\overline{q}g_1g_2$. 
The result is shown as a function of the cutoff 
$y_{cut}$.  The rate decreases slowly, from 0.72 to 0.67, 
as $y_{cut}$ variates from $10^{-4}$ to $10^{-2}$. 
These values are smaller than the rate 0.83 obtained 
by neglecting the kinematic interference 
contribution. The
difference is not large, which implies that kinematic 
interference terms are less important 
than non-interference ones, but
not negligibly small. 
Therefore we may expect that the fraction of 
non-chain states is an estimate of the fraction
of events where color reconnection is possible.

Similar to $n!$ singlet chain 
states, there are also $n!$ terms in the 
color Hamiltonian,
where each term 
corresponds to an order between
$n$ gluons. Up to $O(\frac 1{N_c^2})$ 
we prove, with the help of a diagram 
technique in section III,
that there exists a {\it one-to-one} correspondence between 
a chain state and the term $D^P$ in $H_c$,
with the same order of gluon labels. 
This means that when computing the fraction of a chain 
state $\left| f\right\rangle $ with a specific 
order of gluon connections, up to $O(1/N_c^2)$
we only need to consider the contribution from 
the term in $H_c$ where the 
gluon labels are in the 
same order, and those terms for which the order
of gluons is most close to it.

Finally we give the explicit form for 
the $D$ function and $H_c$ in a special case, 
the case of soft gluon bremsstrahlung. 
In soft gluon bremsstrahlung with gluons 
strongly ordered in energy or transverse momentum, 
the 
emission amplitude has an eikonal form. 
In this case, we have shown that in leading 
order in $N_c$, the cross section of 
$e^{+}e^{-}\rightarrow q\overline{q}+ng$ can be decomposed 
into $n!$ independent or incoherent parts and each part 
represents the contribution from the chain state 
with a specific order of gluons.
We also give the next-to-leading order corrections 
for $e^+e^-\rightarrow q\overline{q}g_1g_2$ 
and $e^+e^-\rightarrow q\overline{q}g_1g_2g_3$. 
The corrections arise from 
interferences of two $D$-functions with 
their gluon orders in the $NL_{1,2}$ or $\overline{NL}$ 
set. Each next-to-leading correction brought by 
the interference of a pair of 
$D$-functions, say, $D^P$ and $D^{P'}$, 
can be regarded as a small perturbation to the dipole sequence 
which corresponds to $|D^P|^2$ or 
$|D^{P'}|^2$ by having dipoles formed by non-adjacent 
gluons, while keeping the rest of the dipole 
sequence same as that corresponding to $|D^P|^2$ or 
$|D^{P'}|^2$. The next-to-leading correction causes the least 
perturbation to the dipole sequence of the leading order 
compared to higher-order corrections.

\vspace{0.5cm}
{\bf Acknowledgment} This work is supported in part by the National 
Natural Science Foundation of China. 

\newpage

\begin{figure} 
 \setlength{\epsfxsize}{3in}
  \centerline{\epsffile{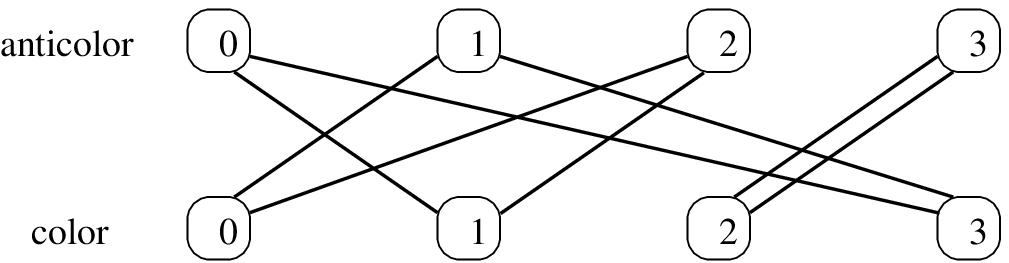}}
\vspace*{.2in}
\caption{The diagram for calculating 
$\langle 1_{q1}1_{12}1_{23}1_{3\overline{q
}}\mid 1_{q2}1_{23}1_{31}1_{1\overline{q}}\rangle $. 
The number of closed
paths is $l=2$. Thus the result is $N_c^2$. }
\label{fig:one}
\vspace*{.4in}
\end{figure}

\begin{figure}
 \setlength{\epsfxsize}{6in}
  \centerline{\epsffile{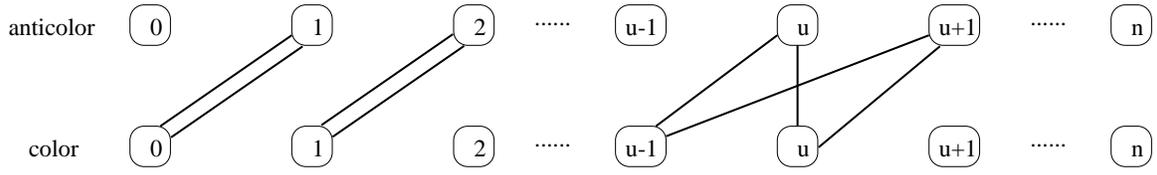}}
\vspace*{.4in}
\caption{The diagram for $\langle 1_{q1}1_{12}1_{23}\cdots 1_{i,i+1}\cdots
1_{n \overline{q}}\mid 1_u1_{q1}1_{23}\cdots 1_{u-2,u-1}1_{u-1,u+1}\cdots
1_{n \overline{q}}\rangle $. There are $(n+1-2)$ double-line loops plus one
loop connecting color $u-1$, $u$ and anticolor $u$, $u+1$.}
\label{fig:two}
\vspace*{.4in}
\end{figure}

\begin{figure}
 \setlength{\epsfxsize}{6in}
  \centerline{\epsffile{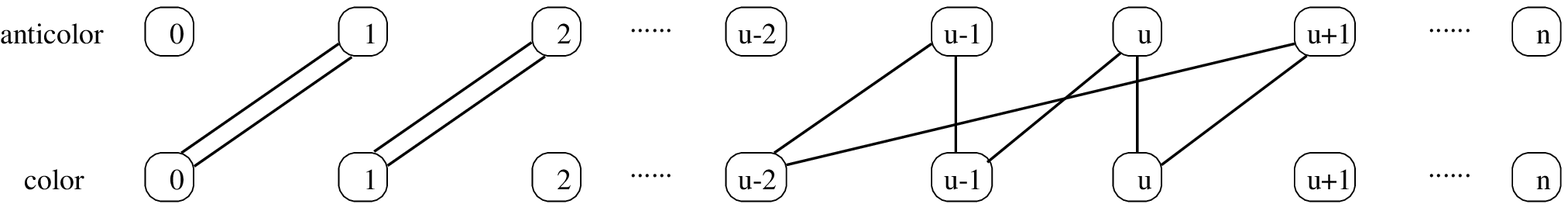}}
\vspace*{.5in}
 \setlength{\epsfxsize}{6in}
  \centerline{\epsffile{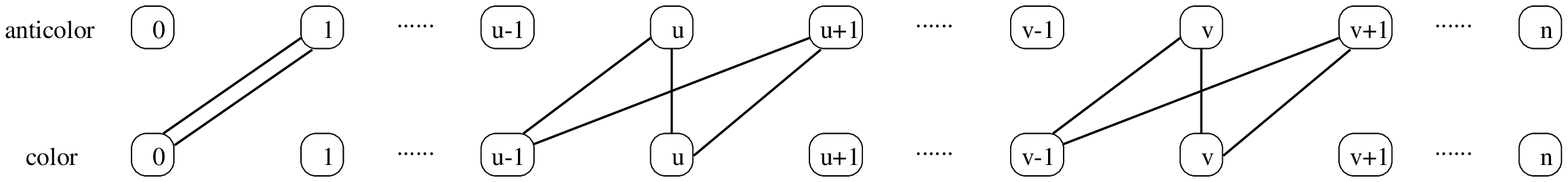}}
\vspace*{.5in}
\caption{The diagram for $\langle 1_{q1}1_{12}\cdots 1_{i,i+1}\cdots 1_{n 
\overline{q}}\mid 1_{u_1}1_{u_2}(1_{qv_1}1_{v_1v_2}\cdots
1_{v_iv_{i+1}}\cdots 1_{v_{n-k}\overline{q}})\rangle $. (a) $u_1$ and $u_2$
are are neighbors. (b) $u_1$ and $u_2$ are not neighbors.
}
\label{fig:three}
\vspace*{.4in}
\end{figure}

\begin{figure}
 \setlength{\epsfxsize}{5in}
  \centerline{\epsffile{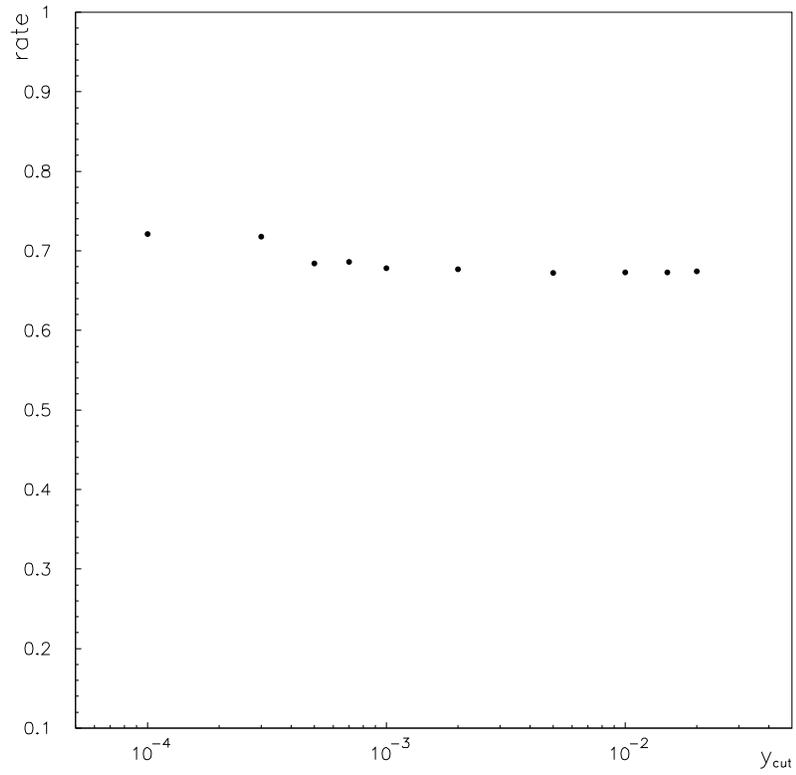}}
\vspace*{.5in}
\caption{The numerical result for the fraction of chain states
in $e^+e^-$ annihilation into two gluons. }
 \label{fig:four}
\vspace*{.4in}
\end{figure}

\end{document}